\documentclass[a4paper,twocolumn]{jpsj3}
\usepackage{bm}
\usepackage{graphicx}

\title{Low-Energy Effective Hamiltonian and the Surface States of
Ca$_{\text{3}}$PbO}
\author{Toshikaze \textsc{Kariyado}\thanks{E-mail address: kariyado@hosi.phys.s.u-tokyo.ac.jp} 
and Masao \textsc{Ogata}}
\inst{Department of Physics, the University of Tokyo, Bunkyo, Tokyo
113-0033, Japan}

\abst{
The band structure of Ca$_{\text{3}}$PbO, which
possesses a three-dimensional massive Dirac electron at the Fermi
energy, is investigated in
detail. Analysis of the orbital weight distributions on the bands
obtained in the first-principles calculation reveals that
the bands crossing the Fermi energy originate from the three Pb-$p$ orbitals
and three Ca-$d_{x^2-y^2}$ orbitals. Taking these Pb-$p$ and
Ca-$d_{x^2-y^2}$ orbitals as basis wave functions, a tight-binding model
is constructed. With the appropriate choice of the hopping integrals and
the strength of the spin--orbit coupling, the constructed model
sucessfully captures important features of the band structure around the
Fermi energy
obtained in the first-principles calculation. By applying the
suitable basis transformation and expanding the matrix elements in the
series of the momentum measured from a Dirac point, the low-energy
effective Hamiltonian of this model is explicitely derived and proved
to be a Dirac Hamiltonain. The origin of the mass term is also
discussed. It is shown that the spin--orbit coupling and the orbitals
other than Pb-$p$ and Ca-$d_{x^2-y^2}$ orbitals play important roles in
making the mass term finite. Finally, the surface band structures of
Ca$_3$PbO for several types of surfaces are investigated using the
constructed
tight-binding model. We find that there appear nontrivial surface states
that cannot be explained as the bulk bands projected on the surface
Brillouin zone. The relation to the topological insulator is also
discussed. 
}

\kword{Dirac electron, inverse perovskite, Ca$_{\text{3}}$PbO, the
first-principles calculation, tight-binding model}

\begin{document}
\maketitle

\section{Introduction}\label{sec1}
A low-energy effective Hamiltonian of a material sometimes becomes
a relativistic Dirac Hamiltonian despite the fact that electrons in
materials are basically described by a nonrelativistic Hamiltonian and
the perturbative treatment of the relativistic effects is generally
legitimated. Such an emergent ``Dirac electron'' is known to host many
intriguing phenomena. The well-known and well studied realizations of the
emergent Dirac electrons in materials are
graphene\cite{PhysRev.71.622,Novoselov:2005fk,RevModPhys.81.109}
(two-dimensional,
massless) and bismuth\cite{Wolff19641057} (three-dimensional,
massive). It is also known that an organic compound,
$\alpha$-(BEDT-TTF)$_{\text{2}}$I$_{\text{3}}$, has Dirac electrons in
the band structure near the Fermi
energy\cite{JPSJ.75.054705,alpha_review}. Rather new example of
emergent Dirac electrons are surface states of three-dimensional topological
insulators\cite{RevModPhys.82.3045}, which
attract great interests and are extensively studied in these
days. Another new example is SrMnBi$_{\text{2}}$, which is claimed to
have a two-dimensional strongly-anisotropic massive Dirac
electron\cite{PhysRevB.84.064428,PhysRevLett.107.126402,PhysRevB.84.220401}.
A Weyl fermion, which has a linear dispersion but is described by
2$\times$2 (not 4$\times$4) matrix is also claimed to exist in
Y$_{\text{2}}$Ir$_{\text{2}}$O$_{\text{7}}$\cite{PhysRevB.83.205101},
where the time reversal symmetry is broken by a magnetic order.

Recently, we proposed an inverse-perovskite material Ca$_{\text{3}}$PbO
and its family as candidates for new materials having Dirac
electrons. The first-principles calculation shows that a
three-dimensional massive Dirac electron appears in the low-energy band
structure of Ca$_{\text{3}}$PbO\cite{JPSJ.80.083704}. A Dirac point,
which is defined as the center of linear dispersion of the Dirac
electron, is located on the $\mathrm{\Gamma}$--X line. Due to the
cubic symmetry of this material, existence of a Dirac
point on the $\mathrm{\Gamma}$--X line implies that there are six equivalent
Dirac points in the Brillouin zone. The most remarkable feature of
the Dirac electron in Ca$_{\text{3}}$PbO is that it appears exactly at the
Fermi energy, and the bands forming the Dirac electron are the only
bands that cross the Fermi energy. Namely, the Dirac type linear dispersion is
the only feature comes up in the vicinity of the Fermi energy. 

This
point gives a strong contrast to bismuth, which is a well-known and well
studied
example having three-dimensional massive Dirac electron in its
band structure. In bismuth, not only bands with Dirac type linear
dispersion, but also a band with parabolic dispersion intersects the Fermi
energy and gives a usual hole Fermi
surface\cite{PhysRevB.41.11827,PhysRevB.52.1566}. As a result, one has to
dope some carriers in order to investigate the properties related to the Dirac
electrons\cite{wehrli}. With this respect, it is important to study
Ca$_{\text{3}}$PbO, which has a simpler band structure than bismuth, in
order to obtain a deeper understanding of three-dimensional Dirac
electrons in materials. 
Furthermore, high symmetry of the crystal structure of
Ca$_3$PbO enables us to have an intuitive view on the origin of 
Dirac electron in
Ca$_3$PbO by analyzing the symmetry of wave functions for the states
relevant to the Dirac electron\cite{JPSJ.80.083704}. The main topic of
this paper is to show the details of such an analysis that was briefly
reported before\cite{JPSJ.80.083704}. The origin of Dirac electron in
bismuth is complicated\cite{Abrikosov,McClure} and its simple and intuitive
picture is still missing. Therefore, the analysis of Ca$_3$PbO
demonstrated in the following will also give hints for the case of bismuth.

In this paper, a detailed analysis of the band structure of
Ca$_{\text{3}}$PbO is presented. First, we construct a tight-binding
model that
captures important features of the band structure near the Fermi
energy obtained in the
first-principles calculation. It is assigned that the
three Pb-$p$ orbitals and three Ca-$d_{x^2-y^2}$ orbitals
[Fig.~\ref{Fig3}(b)] should be included in the tight-binding model by
analyzing the orbital weight distributions on the bands. With the proper
choice of the hopping parameters and the strength of the spin--orbit
coupling, the band structure obtained in the first-principles
calculation is fairly well reproduced by the constructed simple
tight-binding model. The analysis is farther proceeded and we prove
that the low-energy effective Hamiltonian is really a Dirac Hamiltonian
by applying a proper basis transformation and by expanding the matrix
elements with respect to the momentum measured from one of
the Dirac points.
It is worth noting that not only the Hamiltonian, but also the basis
wave functions of the low-energy effective model are explicitly
obtained. These basis wave functions play a role of pseudospins of the
Dirac model. We will also discuss the mass term of
the Dirac Hamiltonian in this material, which has been only briefly
mentioned in our previous paper\cite{JPSJ.80.083704}. Especially, the
roles of the spin--orbit coupling and the orbitals other than Pb-$p$ and 
Ca-$d_{x^2-y^2}$ orbitals are explained. The relation between
Ca$_3$PbO and a topological insulator is also discussed. It is shown
that Ca$_3$PbO is not a topological insulator. Finally, the surface band
structures of Ca$_3$PbO are studied using the constructed
tight-binding model. It is found that there exist nontrivial surface
bands that are nondegenerate and are not explained as the bulk
states projected on the surface Brillouin zone. 

This paper is organized as follows. Section~\ref{sec2}
is used to
describe details of the method for calculation. In \S\ref{sec3}, the
orbital weight distributions on the obtained bands are analyzed and a
tight-binding model is constructed. Section~\ref{sec4} is devoted for
analyzing the obtained tight-binding model and proving that its
low-energy part is actually described by a Dirac Hamiltonian. At the
end of \S\ref{sec4}, the origin of the mass term is also
discussed. In \S\ref{sec:surface}, we make arguments on the topological
properties and the surface band structures.
The paper is summarized in \S\ref{sec6}.

\section{First-Principles Calculation}\label{sec2}
\begin{figure}[htbp]
 \begin{center}
  \includegraphics[scale=1.0]{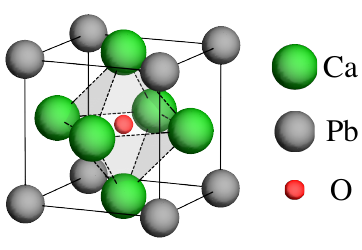}
  \caption{(Color online) Crystal structure of
  Ca$_{\text{3}}$PbO.}\label{Fig1}
 \end{center}
\end{figure}

The band structure calculation is carried out using WIEN2k
package\cite{wien2k}, in which the full-potential augmented-plane-wave
method is implemented. The crystal parameters required in the
calculation are taken from the experimental
results\cite{Widera19801805}. Figure~\ref{Fig1} shows the crystal
structure of Ca$_{\text{3}}$PbO.
Perdew-Burke-Ernzerhof generalized-gradient
approximation\cite{PhysRevLett.77.3865} is used for the exchange-correlation
functional. The parameter choice of ($RK_{\text{max}}$,
$G_{\text{max}}$) $=$ (9.0, 14.0) is used, while radii of spherical atomic
region for Ca, Pb, and O are chosen to
be 2.28, 2.50, and 2.28 (a.u.), respectively. 220 momentum
points in the irreducible Brillouin zone, which are reduced from
20$\times$20$\times$20 momentum points in the full Brillouin zone, are
employed in the self-consistent cycle in the present calculation.
A spin--orbit coupling is taken into account within the spherical atomic
region of each atom via second variational step\cite{0022-3719-13-14-009}. 

We have verified that changes in the parameters stated above
($RK_{\text{max}}$, $G_{\text{max}}$, radii of spherical atomic region,
and the number of momentum points) do
not modify the results in this paper, e.g., appearance of the Dirac type
dispersion in the vicinity of Fermi energy. Although the experimental
crystal parameters are used in the
calculation, it should be noted that the previous theoretical works
showed that the
optimized lattice constants for Ca$_{\text{3}}$PbO and its family
obtained in the first-principles calculation agree well with the
experimental data\cite{Haddadi20101995,Cherrad20112714}, indicating the
consistency between theory and experiments.

A special care should be paid in the treatment of the
spin--orbit coupling. Namely, we should be careful in applying the
second variational step to heavy elements such as Pb in which the
spin--orbit
coupling is expected to be strong\cite{0022-3719-13-14-009}.
However, we think that this is not a serious problem in our
calculation. One of the reasons is that only the state with the total
angular momentum $J=3/2$ is important for the Dirac electron in this
material as will be explained later, while the second variational
step mainly causes problems for $J=1/2$
state\cite{0022-3719-13-14-009}. Another reason is that
the band structure does not change (except for unimportant points) even
if we use the pseudopotential method\cite{QE-2009} with fully-relativistic
pseudopotential for Pb atom in which the spin--orbit coupling is treated
better than in the second-variational step.

\section{Construction of the Tight-Binding Model}\label{sec3}
\begin{figure*}[tb]
 \begin{center}
  \includegraphics[scale=1.0]{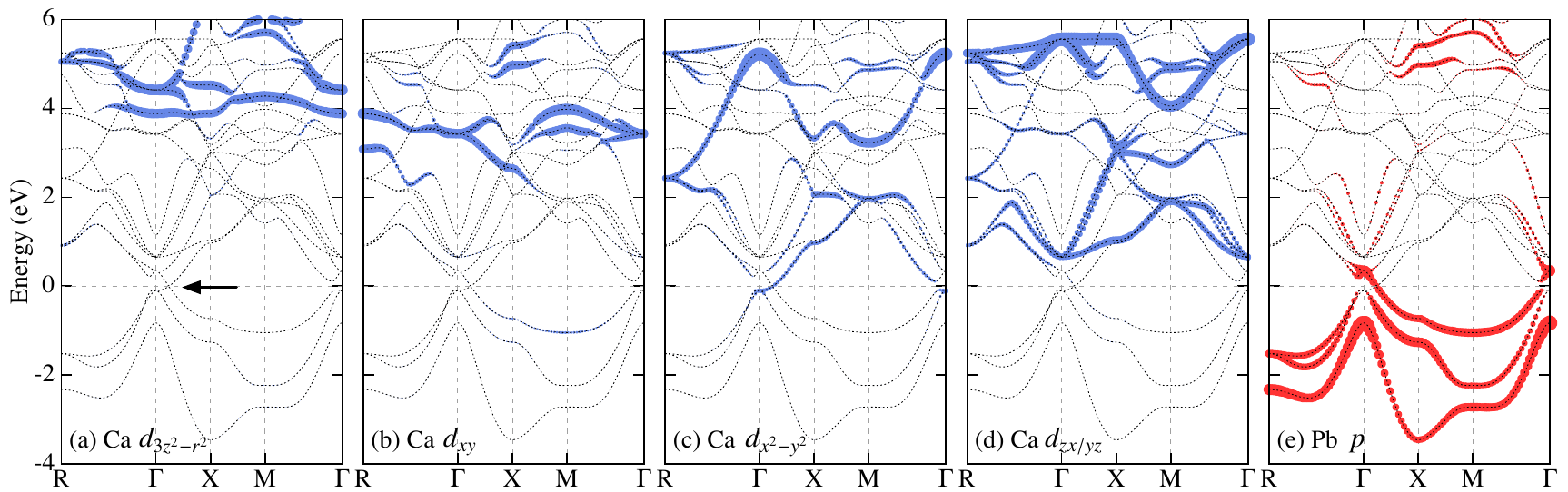}
  \caption{(Color online) Orbital-weight distributions on the
  bands obtained in the first-principles calculation. (a)--(e) show the
  distribution of the Ca $d_{3z^2-r^2}$,
  Ca $d_{xy}$, Ca $d_{x^2-y^2}$, Ca $d_{zx/yz}$, and Pb $p$ orbital
  weights, respectively. The orbital weights are represented as the
  width of the corresponding bands.}
  \label{Fig2}
 \end{center}
\end{figure*}

\subsection{Orbital Character of Each Band}
The obtained band structure of Ca$_{\text{3}}$PbO is presented in
Fig.~\ref{Fig2} together with the orbital weight distributions for Ca
$d_{3z^2-r^2}$, $d_{xy}$, $d_{x^2-y^2}$, $d_{zx/yz}$, and Pb
$p$ orbitals. As is
shown in ref.~\citen{JPSJ.80.083704}, there appears a Dirac electron on
the $\mathrm{\Gamma}$--X line, and the Dirac point is marked by an arrow
in Fig.~\ref{Fig2}(a). The appearance of a Dirac point on the
$\mathrm{\Gamma}$--X line implies that there are six Dirac points
in the whole Brillouin zone due to the cubic symmetry of this material.
In other words, Dirac points are found at $(k_0,0,0)$,
$(-k_0,0,0)$, $(0,k_0,0)$, $(0,-k_0,0)$, $(0,0,k_0)$, and
$(0,0,-k_0)$. Although it is difficult to see in the presented energy scale,
there exists a very small gap at the Fermi energy, and the emerged Dirac
electron is actually massive with a very small mass. The magnitude of
the mass gap is about 15 meV\cite{JPSJ.80.083704}.

\begin{figure}[htbp]
 \begin{center}
  \includegraphics[scale=1.0]{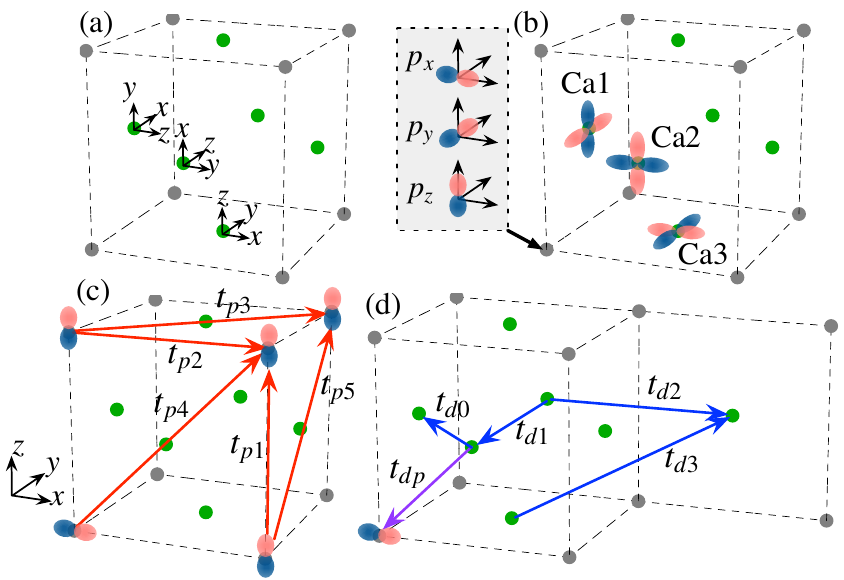}
  \caption{(Color online) (a) The local coordinates on the three Ca
  sites in a unit cell. (b) Schematic pictures of the basis wave
  functions used in the reduced tight-binding model.  (c,d)
  Definitions of hopping integrals between the two $p$-orbitals
  (c), between the two $d$-orbitals and between the $p$- and
  $d$-orbitals (d).}
  \label{Fig3}
 \end{center}
\end{figure}

Before moving on to the discussion on the orbital
weight distributions, we explain the symmetry of the local environment
of each atom. The local environment of the Ca atom has a tetragonal
symmetry with its tetragonal axis directing along a line connecting the Ca
and O atoms (see Fig.~\ref{Fig1}). The tetragonal-axis directions of
the three Ca sites in the unit cell are different from each other. Then,
it is convenient to introduce local coordinates for each Ca atom,
whose definitions are illustrated in Fig.~\ref{Fig3}(a). Using these
local coordinates, the Ca-$3d$ orbitals can be classified into four
groups, i.e., $d_{3z^2-r^2}$, $d_{xy}$, $d_{x^2-y^2}$, and $d_{zx/yz}$,
reflecting the local tetragonal symmetry. Note that $x$, $y$, and $z$
appearing in the subscripts refer to the local coordinates. On the other
hand, the
local environment of the Pb atom has a cubic symmetry and all Pb-$6p$
orbitals ($p_x$, $p_y$, and $p_z$) are classified into a single group
named as Pb-$p$. 

The orbital weight distributions of Ca-$d_{3z^2-r^2}$, $d_{xy}$,
$d_{x^2-y^2}$, and $d_{zx/yz}$ orbitals and Pb-$p$ orbitals are plotted
in Figs.~\ref{Fig2}(a)--\ref{Fig2}(e), where the weights are indicated
by the width of each band. We can see that in the presented energy
range, the bands below the Fermi energy mainly originate from Pb-$p$
orbitals,
while the highly entangled bands above the Fermi energy mainly originate
from Ca-$3d$ orbitals. However, note that the top of p-bands (bands
originate from Pb-$6p$ orbitals) locates above the bottom of d-bands
(bands originate from Ca-$3d$ orbitals) [Figs.~\ref{Fig2}(c) and
\ref{Fig2}(e)]. As will be discussed later, this overlap between the p-
and d-bands is essential for the emergence of Dirac electron.
 Among the Ca-$3d$ orbitals, the ``center of
mass'' of Ca-$d_{3z^2-r^2}$ orbital weights lies at relatively high
energy while that of Ca-$d_{x^2-y^2}$ lies at relatively
low energy. This tendency can be understood from the crystal
field splitting. Specifically, the orbitals whose robes are directed to O or
Pb (like Ca-$d_{3z^2-r^2}$) tend to have higher energy than the other
orbitals since O and Pb are
negatively charged in this material. On the other hand, the orbitals
whose robes are directed to direction other than O or Pb (like
Ca-$d_{x^2-y^2}$) tend to have low energy.
Note that Ca-$d_{x^2-y^2}$
and Ca-$d_{zx/yz}$ orbitals have large dispersion and their band widths
exceed the energy difference of the center of masses of each orbital.

The most important information obtained from
Figs.~\ref{Fig2}(a)--\ref{Fig2}(e) is that two bands
forming the Dirac electron near the Fermi energy come from
Ca-$d_{x^2-y^2}$ and Pb-$p$ orbitals [see
Figs.~\ref{Fig2}(c) and \ref{Fig2}(e)]. For Ca-$d_{x^2-y^2}$ orbital,
there are two reasons for this orbital to come down to the Fermi energy:
one is its
large dispersion and the other is that its center of mass lies at
relatively low energy. On the basis of this observation, 
we neglect other Ca-$d$ orbitals ($d_{3z^2-r^2}$, $d_{xy}$, and $d_{zx/yz}$)
in constructing a model describing the Dirac electron in this
material. 

\subsection{Tight-Binding Model}
Now, we construct a tight-binding model that describes the
low-energy band structure of Ca$_{\text{3}}$PbO. Based on the
observations in the previous subsection, six orbitals,
namely, three Pb-$p$ orbitals ($p_x$, $p_y$, and $p_z$) and three
Ca-$d_{x^2-y^2}$ orbitals in the unit cell are taken as basis
orbitals. The three
Ca-$d_{x^2-y^2}$ orbitals locate on Ca1, Ca2, and Ca3 site,
respectively. See Figs.~\ref{Fig1} and \ref{Fig3}(b) for the definitions
of Ca1, Ca2, and Ca3 sites and the schematic pictures for the basis
orbitals. Note that $x$, $y$, and $z$ appearing in the subscripts refer to
the local coordinates as in Figs.~\ref{Fig3}(a) and \ref{Fig3}(b).
Furthermore, due to the strong spin--orbit coupling on
the Pb atom, the spin degrees of freedom should be explicitly
treated. As a result, we use 12 ($=6\times 2$) orbitals
\begin{subequations}
\begin{align}
 |p_x\uparrow\rangle, && |p_y\uparrow\rangle, && |p_z\uparrow\rangle, &&
 |p_x\downarrow\rangle, && |p_y\downarrow\rangle, &&
 |p_z\downarrow\rangle,\label{basis_p}\\
 |d_1\uparrow\rangle, &&  |d_2\uparrow\rangle, &&  |d_3\uparrow\rangle, && 
 |d_1\downarrow\rangle, &&  |d_2\downarrow\rangle, &&  |d_3\downarrow\rangle.\label{basis_d}
\end{align}\label{basis_original}
\end{subequations}

Using these local orbitals as the basis set, we construct a
tight-binding Hamiltonian that is written as 
\begin{multline}
  \hat{H}
  =\sum_\sigma\sum_{\bm{r}a\bm{r}'a'}
  t_{aa'}(\bm{r}-\bm{r}')
  c^\dagger_{\bm{r}a\sigma}c_{\bm{r}'a'\sigma}\\
 +\sum_{\bm{r}a\sigma\bm{r'}a'\sigma'}
 \lambda^{\sigma\sigma'}_{aa'}(\bm{r}-\bm{r}')
 c^\dagger_{\bm{r}a\sigma}c_{\bm{r}'a'\sigma'}
  \label{Hamiltonian0}
\end{multline}
where the indices $a$ and $a'$ represent $p_{x,y,z}$ or
$d_{1,2,3}$, respectively. Hopping processes described in
Figs.~\ref{Fig3}(c) and \ref{Fig3}(d) are included in
the first term of eq.~\eqref{Hamiltonian0} (details are explained soon
later). The second term represents the spin--orbit coupling, which is 
assumed here to act only on the Pb-$p$ orbitals for simplicity. The
spin--orbit coupling on the Ca-$d$ orbitals is neglected. The Fourier
transformation of eq.~\eqref{Hamiltonian0} gives
\begin{equation}
  \hat{H}
  =\sum_{\bm{k}}\sum_{\alpha\alpha'}\mathcal{E}_{\alpha\alpha'}(\bm{k})
  c^\dagger_{\bm{k}\alpha}c_{\bm{k}\alpha'},
\end{equation}
where the indices $\alpha$ and $\alpha'$ run through 1 to 12,
corresponding to the 12
basis orbitals in the order written in eqs.~\eqref{basis_p} and
\eqref{basis_d}. 
As has been carried out in ref.~\citen{JPSJ.80.083704}, the matrix elements are
transformed by attaching momentum-dependent phase factors to the basis
orbitals as 
$|p_{x,y,z}\sigma\rangle\rightarrow\mathrm{e}^{\mathrm{i}(k_x+k_y+k_z)/2}|p_{x,y,z}\sigma\rangle$, 
$|d_1\sigma\rangle\rightarrow\mathrm{e}^{\mathrm{i}k_x/2}|d_1\sigma\rangle$,
$|d_2\sigma\rangle\rightarrow\mathrm{e}^{\mathrm{i}k_y/2}|d_2\sigma\rangle$, 
and
$|d_3\sigma\rangle\rightarrow\mathrm{e}^{\mathrm{i}k_z/2}|d_3\sigma\rangle$. 
This is performed for the matrix elements to have simple forms. 
The transformed basis and matrix elements are used hereafter.

Now, the Hamiltonian can be written in a compact form as 
\begin{equation}
 \hat{\mathcal{E}}_{\bm{k}}=
  \begin{pmatrix}
   \hat{\tilde{\mathcal{E}}}^{pp}_{\bm{k}}+\hat{\lambda}_{\bm{k}}&
   \hat{\tilde{\mathcal{E}}}^{pd}_{\bm{k}}\\
   (\hat{\tilde{\mathcal{E}}}^{pd}_{\bm{k}})^\dagger&
   \hat{\tilde{\mathcal{E}}}^{dd}_{\bm{k}}
  \end{pmatrix},
\end{equation}
where
\begin{equation}
 \hat{\tilde{\mathcal{E}}}_{\bm{k}}^{qq'}=
  \begin{pmatrix}
   \hat{\mathcal{E}}_{\bm{k}}^{qq'}&\hat{0}\\
   \hat{0}&\hat{\mathcal{E}}_{\bm{k}}^{qq'}
  \end{pmatrix}
  \quad (q,q'=p\text{ or }d)\label{Etilde}
\end{equation}
and
\begin{equation}
 \hat{\lambda}_{\bm{k}}=\hat{\lambda}^{(0)}+\hat{\lambda}^{(1)}_{\bm{k}}. 
\end{equation}
Here, $\hat{\tilde{\mathcal{E}}}^{qq'}_{\bm{k}}$ and
$\hat{\lambda}_{\bm{k}}$ are 6$\times$6 matrices, while
$\hat{\mathcal{E}}^{qq'}_{\bm{k}}$ is a 3$\times$3
matrix. $\hat{\tilde{\mathcal{E}}}^{qq'}_{\bm{k}}$ ($q,q'=p$ or $d$)
represents usual hopping processes. $\hat{0}$ in eq.~\eqref{Etilde} is
due to the fact that the nonrelativistic Hamiltonian has no terms mixing
the spin components. $\hat{\lambda}_{\bm{k}}$
represents the spin--orbit
coupling that is caused by relativistic effects, and has matrix elements
mixing the spin
components. In the following subsections, we explain details of each
matrix in order. 

\subsubsection{Spin--Orbit Coupling}
First, $\hat{\lambda}_{\bm{k}}$ contains the contributions from the spin--orbit
coupling on the Pb-$p$ orbitals. $\hat{\lambda}^{(0)}$
represents the on-site $\bm{L}\cdot\bm{S}$ coupling for the $p$-orbitals
and is explicitly written as
\begin{equation}
 \hat{\lambda}^{(0)}=\lambda_0
  \begin{pmatrix}
   0&-\mathrm{i}&0&0&0&1\\
   \mathrm{i}&0 &0 &0 &0 &-\mathrm{i}\\ 
   0&0 &0 &-1 &\mathrm{i} &0\\ 
   0&0 &-1 &0 &\mathrm{i} &0\\ 
   0&0 &-\mathrm{i} &-\mathrm{i} &0 &0\\ 
   1&\mathrm{i} &0 &0 &0 &0 
  \end{pmatrix}.
\end{equation}
Next, $\hat{\lambda}^{(1)}_{\bm{k}}$ represents the
intersite counterpart of the $\bm{L}\cdot\bm{S}$ coupling and is
explicitly written as
\begin{equation}
 \hat{\lambda}^{(1)}_{\bm{k}}=\lambda_1
  \begin{pmatrix}
   0&-\mathrm{i}\bar{c}_z&0&0&0&\bar{c}_y\\
   \mathrm{i}\bar{c}_z&0 &0 &0 &0 &-\mathrm{i}\bar{c}_x\\ 
   0&0 &0 &-\bar{c}_y &\mathrm{i}\bar{c}_x &0\\ 
   0&0 &-\bar{c}_y &0 &\mathrm{i}\bar{c}_z &0\\ 
   0&0 &-\mathrm{i}\bar{c}_x &-\mathrm{i}\bar{c}_z &0 &0\\ 
   \bar{c}_y&\mathrm{i}\bar{c}_x &0 &0 &0 &0 
  \end{pmatrix},\label{lambda1}
\end{equation}
where $\bar{c}_x=c_y+c_z$, $\bar{c}_y=c_z+c_x$, and $\bar{c}_z=c_x+c_y$
with $c_a=\cos k_a$ ($a=x,y,z$). 
Here the intersite coupling means the coupling between the nearest
neighbor pair of two Pb-$p$ orbitals. (Ca-$d$ orbitals are not involved
in this coupling.) $\hat{\lambda}^{(1)}_{\bm{k}}$ was neglected in
ref.~\citen{JPSJ.80.083704} since only $\hat{\lambda}^{(0)}$ plays an
important role in reproducing the main
features of the band structure obtained in the first-principles
calculation. However, a close observation of the band structure 
reveals that the energy splittings
between the states with $J=1/2$ and $J=3/2$ take different values
at the $\mathrm{\Gamma}$- and at the R-point, which can be
explained only if we take account of $\hat{\lambda}^{(1)}_{\bm{k}}$. In
the following formulation, we keep $\hat{\lambda}^{(1)}_{\bm{k}}$ in
order to see its effect, although the actual magnitude of $\lambda_1$ in
eq.~\eqref{lambda1} is small.

\subsubsection{$p$-$p$ matrix elements}
Next, $\hat{\mathcal{E}}^{pp}_{\bm{k}}$ contains the on-site energy of
the Pb-$p$ orbital and hoppings between the two Pb-$p$
orbitals. Considering all the hopping processes between the
nearest- and the next-nearest-neighbor pairs of Pb-$p$ orbitals as shown
in Fig.~\ref{Fig3}(c), $\hat{\mathcal{E}}^{pp}_{\bm{k}}$ becomes
\begin{equation}
  \hat{\mathcal{E}}^{pp}_{\bm{k}}=
   \begin{pmatrix}
    \epsilon_{xx}&\epsilon_{xy}&\epsilon_{xz}\\
    \epsilon_{yx}&\epsilon_{yy}&\epsilon_{yz}\\
    \epsilon_{zx}&\epsilon_{zy}&\epsilon_{zz}
   \end{pmatrix},
\end{equation}
where
\begin{subequations} 
\begin{align}
 \epsilon_{xx}
 &=\epsilon_p+2t_{p1}c_x+2t_{p2}\bar{c}_x+4t_{p3}c_yc_z+4t_{p5}c_x\bar{c}_x,\\
 \epsilon_{yy}
 &=\epsilon_p+2t_{p1}c_y+2t_{p2}\bar{c}_y+4t_{p3}c_zc_x+4t_{p5}c_y\bar{c}_y,\\
 \epsilon_{zz}
 &=\epsilon_p+2t_{p1}c_z+2t_{p2}\bar{c}_z+4t_{p3}c_xc_y+4t_{p5}c_z\bar{c}_z,
\end{align}
\end{subequations}
and $\epsilon_{xy}=-4t_{p4}s_xs_y$, $\epsilon_{yz}=-4t_{p4}s_ys_z$, and
$\epsilon_{zx}=-4t_{p4}s_zs_x$ with $s_a=\sin k_a$ ($a=x,y,z$). Note
that $t_{p4}$ and $t_{p5}$ were not included in ref.~\citen{JPSJ.80.083704}
for simplicity. These terms improve the agreement between the band
structure of the tight-binding model and that of the first-principles
calculation, although the results do not change qualitatively.
\subsubsection{$d$-$d$ matrix elements}
Next, $\hat{\mathcal{E}}^{dd}_{\bm{k}}$ contains the
on-site energy of the Ca-$d$ orbital and hoppings between the two Ca-$d$
orbitals. Considering up to the third-nearest-neighbor pairs of Ca-$d$
orbitals as shown in Fig.~\ref{Fig3}(d),
$\hat{\mathcal{E}}^{dd}_{\bm{k}}$ becomes
\begin{equation}
 \hat{\mathcal{E}}_{\bm{k}}^{dd}=
  \begin{pmatrix}
   \epsilon_{11}&\epsilon_{12}&\epsilon_{13}\\
   \epsilon_{21}&\epsilon_{22}&\epsilon_{23}\\
   \epsilon_{31}&\epsilon_{32}&\epsilon_{33}
  \end{pmatrix},
\end{equation}
where
\begin{subequations} 
 \begin{align}
 \epsilon_{11}&=\epsilon_d+2t_{d1}c_x+2t_{d2}\bar{c}_x,\\
 \epsilon_{22}&=\epsilon_d+2t_{d1}c_y+2t_{d2}\bar{c}_y,\\
 \epsilon_{33}&=\epsilon_d+2t_{d1}c_z+2t_{d2}\bar{c}_z,\\
 \epsilon_{12}&=4(t_{d0}+2t_{d3}c_z)c_{\frac{x}{2}}c_{\frac{y}{2}},\\
 \epsilon_{23}&=4(t_{d0}+2t_{d3}c_x)c_{\frac{y}{2}}c_{\frac{z}{2}},\\
 \epsilon_{31}&=4(t_{d0}+2t_{d3}c_y)c_{\frac{z}{2}}c_{\frac{x}{2}},
 \end{align}
\end{subequations}
with $c_{\frac{a}{2}}=\cos \frac{k_a}{2}$ ($a=x,y,z$). Note that
$t_{d3}$ was not included in ref.~\citen{JPSJ.80.083704} again, and is
included here for completeness and to improve the tight-binding model.
\subsubsection{$d$-$p$ matrix elements}
Finally, $\hat{\mathcal{E}}^{pd}$ represents the
hybridization between the nearest-neighbor pair of the Pb-$p$ and
the Ca-$d$ orbitals, which is written as
 \begin{equation}
 \hat{\mathcal{E}}^{pd}_{\bm{k}}=
  \begin{pmatrix}
   0&\epsilon_{x2}&\epsilon_{x3}\\
   \epsilon_{y1}&0&\epsilon_{y3}\\
   \epsilon_{z1}&\epsilon_{z2}&0
  \end{pmatrix},
 \end{equation}
where 
$\epsilon_{x2}=-4\mathrm{i}t_{dp}s_{\frac{x}{2}}c_{\frac{z}{2}}$,
$\epsilon_{x3}= 4\mathrm{i}t_{dp}s_{\frac{x}{2}}c_{\frac{y}{2}}$,
$\epsilon_{y3}=-4\mathrm{i}t_{dp}s_{\frac{y}{2}}c_{\frac{x}{2}}$,
$\epsilon_{y1}= 4\mathrm{i}t_{dp}s_{\frac{y}{2}}c_{\frac{z}{2}}$,
$\epsilon_{z1}=-4\mathrm{i}t_{dp}s_{\frac{z}{2}}c_{\frac{y}{2}}$, and
$\epsilon_{z2}= 4\mathrm{i}t_{dp}s_{\frac{z}{2}}c_{\frac{x}{2}}$ with
$s_{\frac{a}{2}}=\sin \frac{k_a}{2}$ ($a=x,y,z$). 

\subsection{Band Structure of the Tight-Binding Model}
Parameters in $\hat{\mathcal{E}}^{qq'}_{\bm{k}}$ and
$\hat{\lambda}_{\bm{k}}$ can be determined from the maximally
localized Wannier function (MLWF) constructing
method\cite{PhysRevB.65.035109,Mostofi2008685,Kunes20101888}. 
However, we find that the parameter set obtained using MLWF, which
includes rather long-ranged hopping processes, does not reproduce the
band structure near the Fermi energy if the hopping integrals are
truncated as in Figs.~\ref{Fig3}(c) and \ref{Fig3}(b).
This is because the long-ranged hopping processes in MLWF are important
for Pb-$p$ orbitals since their wave
functions have large spreadings, while the hopping integrals in
Figs.~\ref{Fig3}(c) and \ref{Fig3}(b) are relatively short-ranged.
In order to overcome this problem, we
modify some parameters obtained in MLWF to reproduce the band structure.
This simplification gives no significant problems in the following
arguments of the appearance of the Dirac electron.

\begin{figure}
 \begin{center}
  \includegraphics[scale=1.0]{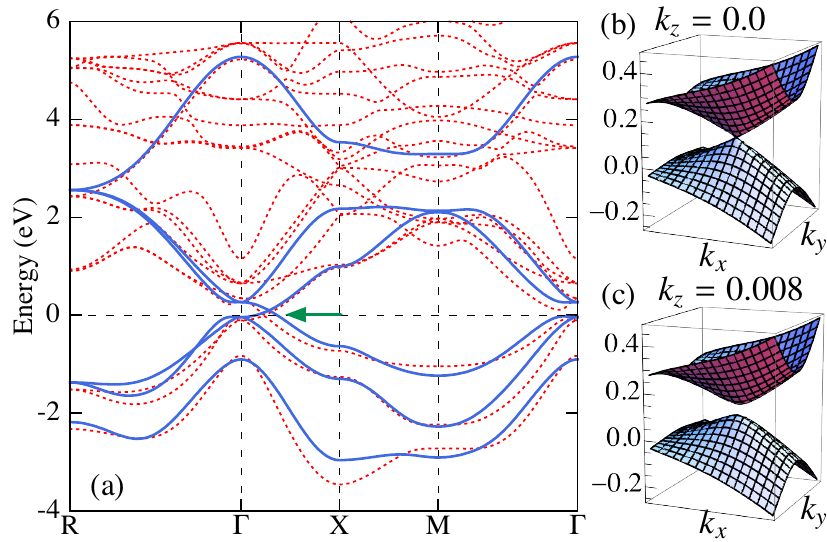}
  \caption{(Color online) (a) Band structure of the tight-binding
  model (solid lines) and that obtained in the first-principles
  calculation (dotted lines). (b,c) Dispersion
  relations of the tight-binding model around a Dirac point. $k_x$ and
  $k_y$ are in the region of $0.06\leq k_x\leq 0.22$ and 
  $-0.08\leq k_y\leq 0.08$, respectively, while $k_z$ is 0 (b), and
  0.008 (c). Momenta are represented in the unit of $2\pi/a$ where $a$
  is the cubic lattice constant.}
  \label{Fig4}
 \end{center}
\end{figure}

The determined parameters are
$\epsilon_p=-1.462$,
$t_{p1}=0.290$, $t_{p2}=0.045$, $t_{p3}=0.035$, $t_{p4}=0.0791$,
$t_{p5}=0.055$, $\epsilon_d=2.146$,
$t_{d0}=0.349$, $t_{d1}=0.184$, $t_{d2}=-0.195$, $t_{d3}=0.047$,
$t_{dp}=0.302$, 
$\lambda_0=0.330$, and $\lambda_1=0.015$ (the units are eV). 
Figure~\ref{Fig4}(a) illustrates the band structure in this
tight-binding model compared with the results of the first-principles
calculation. Comparing
Fig.~\ref{Fig4}(a) and Figs.~\ref{Fig2}(c) and \ref{Fig2}(e), we can see
that the tight-binding model well reproduces the bands with large
Ca-$d_{x^2-y^2}$ [Fig.~\ref{Fig2}(c)] and Pb-$p$ [Fig.~\ref{Fig2}(e)]
orbital weights. This is natural since we have taken account of these
orbitals in
constructing the tight-binding model. Note that not only the global band
structure, but also the most important feature of this system, i.e.,
the Dirac type linear dispersion in the vicinity of the Fermi energy, is
captured.

In order to
check this point, Figs.~\ref{Fig4}(b) and \ref{Fig4}(c) provide
three-dimensional plots of the dispersion
relations of the tight-binding model around one of the Dirac points,
$(k_0,0,0)$. Figure~\ref{Fig4}(b) for $k_z=0$ shows that two cone-shaped
bands touch with each other at an isolated point in the $k_x$--$k_y$
plane, while Fig.~\ref{Fig4}(c) shows that a gap emerges between the two
bands as $k_z$ becomes finite. These observations confirm the
existence of a three-dimensional Dirac electron in this
model. However, we find that the Dirac electron in this
tight-binding model is {\it massless} while the first-principles
calculation gives a small mass gap of 15 meV. We find that this mass
term originates
from the contribution from the other orbitals which are not included in
the present tight-binding model. This will be discussed later in
\S\ref{sec5}. 

\section{Low-Energy Effective Hamiltonian}\label{sec4}
In this section, we concentrate on the Dirac point on the $k_z$-axis,
namely, the Dirac point at $(0,0,k_0)$. This is because we are familiar
with choosing the $z$-axis as the quantization axis of the spin. When we 
discuss the band structure around other Dirac points, such as
$(k_0,0,0)$, it is convenient to direct the spin quantization axis along
the $x$-axis.
\subsection{Analysis of the Tight-Binding Model}
\begin{figure}[htbp]
 \begin{center}
  \includegraphics[scale=1.0]{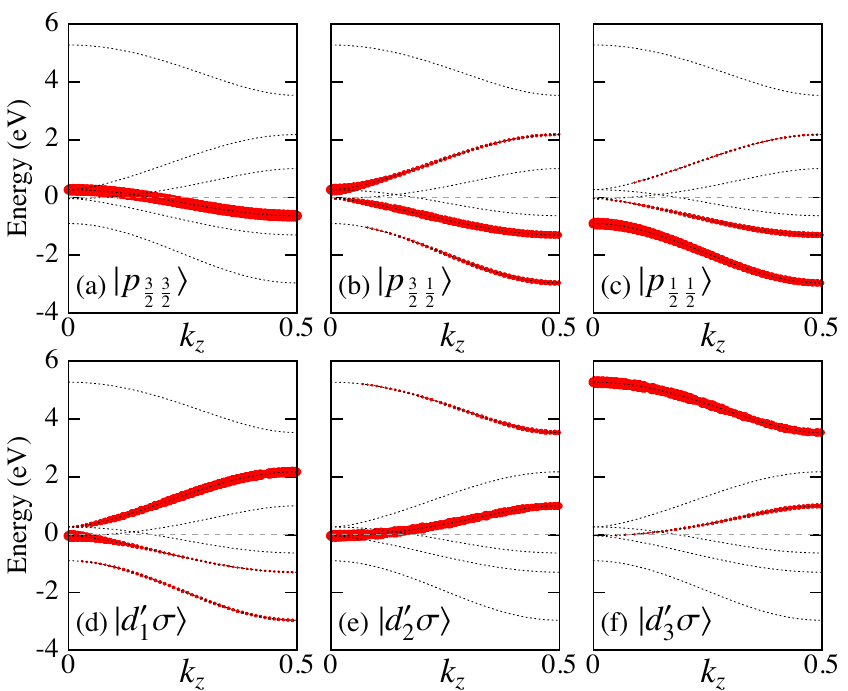}
  \caption{(Color online) Weight distributions of the new basis states on the
  bands along the $k_z$-axis. (a)--(f) show the weights of
  $|p_{\frac{3}{2}\frac{3}{2}}\rangle$, 
  $|p_{\frac{3}{2}\frac{1}{2}}\rangle$, 
  $|p_{\frac{1}{2}\frac{1}{2}}\rangle$, 
  $|d'_1\sigma\rangle$, 
  $|d'_2\sigma\rangle$, and 
  $|d'_3\sigma\rangle$, respectively. 
  }
  \label{Fig5}
 \end{center}
\end{figure}
In order to prove that the low-energy effective Hamiltonian of this
material is really a Dirac Hamiltonian, we first introduce a new basis
set. Here, the eigenstates at the $\mathrm{\Gamma}$-point
are taken as the new basis set since the Dirac point under consideration
is relatively close to the $\mathrm{\Gamma}$-point. In addition, the high
symmetry of the $\mathrm{\Gamma}$-point allows us to classify the eigenstates
easily. Explicitly, the new basis wave functions are defined as
\begin{subequations} 
 \begin{align}
 |p_{\frac{3}{2}\frac{3}{2}}\rangle&=
 -\frac{1}{\sqrt{2}}\bigl(|p_x\uparrow\rangle+\mathrm{i}|p_y\uparrow\rangle\bigr),\\
 |p_{\frac{3}{2}\frac{1}{2}}\rangle&=
 -\frac{1}{\sqrt{6}}\bigl(|p_x\downarrow\rangle+\mathrm{i}|p_y\downarrow\rangle-2|p_z\uparrow\rangle\bigr),\\
  |p_{\frac{3}{2}\bar{\frac{1}{2}}}\rangle&=
  \frac{1}{\sqrt{6}}
   \bigl(|p_x\uparrow\rangle-\mathrm{i}|p_y\uparrow\rangle+2|p_z\downarrow\rangle\bigr),\\
  |p_{\frac{3}{2}\bar{\frac{3}{2}}}\rangle&=
  \frac{1}{\sqrt{2}}\bigl(|p_x\downarrow\rangle-\mathrm{i}|p_y\downarrow\rangle\bigr),\\
  |p_{\frac{1}{2}\frac{1}{2}}\rangle&=
  \frac{1}{\sqrt{3}}
   \bigl(|p_x\downarrow\rangle+\mathrm{i}|p_y\downarrow\rangle+|p_z\uparrow\rangle\bigr),\\
  |p_{\frac{1}{2}\bar{\frac{1}{2}}}\rangle&=
  \frac{1}{\sqrt{3}}
  \bigl(|p_x\uparrow\rangle-\mathrm{i}|p_y\uparrow\rangle-|p_z\downarrow\rangle\bigr),
 \end{align}\label{new_basis_p}\end{subequations}
 and
\begin{subequations}
 \begin{align}
  |d_1'\sigma\rangle&=\frac{1}{\sqrt{2}}
  \bigl(|d_1\sigma\rangle-|d_2\sigma\rangle\bigr),\\
  |d_2'\sigma\rangle&=\frac{1}{\sqrt{6}}
  \bigl(|d_1\sigma\rangle+|d_2\sigma\rangle-2|d_3\sigma\rangle\bigr),\\
  |d_3'\sigma\rangle&=\frac{1}{\sqrt{3}}
  \bigl(|d_1\sigma\rangle+|d_2\sigma\rangle+|d_3\sigma\rangle\bigr).
 \end{align}\label{new_basis_d}\end{subequations}
Here, $|p_{JJ_z}\rangle$ is just the wave function having a total
angular momentum $J$ with its $z$-component $J_z$ realized from the
on-site spin--orbit coupling of the Pb-$p$ orbitals. The bases
$|d'_1\sigma\rangle$, $|d'_2\sigma\rangle$, and $|d'_3\sigma\rangle$
originate from the Ca-$d$
orbitals [eq.~\eqref{new_basis_d}], and spin up and spin
down components are not mixed since the spin--orbit coupling on the
Ca atoms is neglected in this model. 

Figure~\ref{Fig5} shows the weight distribution of the new basis
states on the bands along the $k_z$-axis. From
this figure, it is clearly seen that the bands crossing
the Fermi energy and forming the Dirac electron come from the 
$|p_{\frac{3}{2}\frac{3}{2}}\rangle$ and $|d'_2\sigma\rangle$
states. Therefore, it will be enough to keep only the four states 
$|p_{\frac{3}{2}\frac{3}{2}}\rangle$,
$|p_{\frac{3}{2}\bar{\frac{3}{2}}}\rangle$, $|d'_2\uparrow\rangle$, and
$|d'_2\downarrow\rangle$ out of 12 states in eqs.~\eqref{new_basis_p}
and \eqref{new_basis_d} in deriving the
low-energy effective Hamiltonian. When we examine Fig.~\ref{Fig5} more
closely, we can see that a small weight of $|d'_3\sigma\rangle$ state is
mixed in
the band forming the Dirac electron [Fig.~\ref{Fig5}(f)]. However,
the mixed weight is so small that it gives no harm in the following
arguments.

\subsection{Expansion around Dirac point}
Reflecting the arguments in the previous subsection, we derive the
low-energy
effective Hamiltonian in the following two-step process. In
the first step, the matrix elements of $\hat{\mathcal{E}}^{qq'}_{\bm{k}}$ are
expanded with respect to $k_x$ and $k_y$ up to the first order (remember
that we are concentrating on the Dirac point on the
$k_z$-axis), and at the same time, the basis set in
eq.~\eqref{basis_original} is transformed into that in
eqs.~\eqref{new_basis_p} and \eqref{new_basis_d}.
In the second step,
matrix elements irrelevant to $|p_{\frac{3}{2}\frac{3}{2}}\rangle$,
$|p_{\frac{3}{2}\bar{\frac{3}{2}}}\rangle$, $|d'_2\uparrow\rangle$, and
$|d'_2\downarrow\rangle$ are
dropped. This procedure is slightly different from that used in
ref.~\citen{JPSJ.80.083704}, but the two procedures are essentially the
same. In the following, we make use of the notation
$k_{\pm}=k_x\pm\mathrm{i}k_y$.

In the first step, i.e., expanding the matrix elements
with respect to $k_x$ and $k_y$ (or $k_\pm$), and changing the basis
set,
$\hat{\mathcal{E}}_{\bm{k}}$ is transformed as
\begin{equation}
 \hat{\mathcal{E}}'_{\bm{k}}=
  \begin{pmatrix}
   \hat{\mathcal{E}}'^{pp}_{\bm{k}}& \hat{\mathcal{E}}'^{dp}_{\bm{k}}\\
   (\hat{\mathcal{E}}'^{dp}_{\bm{k}})^\dagger& \hat{\mathcal{E}}'^{dd}_{\bm{k}}
  \end{pmatrix}.\label{Edash}
\end{equation}
Here, $\hat{\mathcal{E}}'^{pp}_{\bm{k}}$ can be
explicitly written as
\begin{equation}
 \hat{\mathcal{E}}'^{pp}_{\bm{k}}=
   \setlength{\arraycolsep}{0.5mm}
  \begin{pmatrix}
   g_1&g_5k_-&0&0&g'_5k_-&0 \\
   g_5k_+&g_2 &0 &0 &g_4 &-g''_5k_- \\
   0&0 &g_2 &-g_5k_- &-g''_5k_+ &-g_4 \\
   0&0 &-g_5k_+ &g_1 &0 & g'_5k_+\\
   g'_5k_+&g_4 &-g''_5k_- &0 &g_3 &0 \\
   0&-g''_5k_+ &-g_4 &g'_5k_- &0 & g_3
  \end{pmatrix}
  \label{Epp}
\end{equation}
with
\begin{subequations} 
 \begin{align}
 g_1&=\lambda_0+2\lambda_1+f_0,\\
 g_2&=\lambda_0+\frac{2}{3}\lambda_1(1+2c_z)+\frac{1}{3}(f_0+2f_1),\\
 g_3&=-2\lambda_0-\frac{4}{3}\lambda_1(2+c_z)+\frac{1}{3}(2f_0+f_1),\\
 g_4&=\frac{\sqrt{2}}{3}\lambda_1(1-c_z)-\frac{\sqrt{2}}{3}(f_0-f_1),\\
 g_5&=4t_{p4}s_z/\sqrt{3},
 \end{align}\label{g1g5}\end{subequations}
and $g'_5=g_5/\sqrt{2}$, $g''_5=\sqrt{3}g'_5$, and 
\begin{subequations} 
  \begin{align}
  f_0&=(\epsilon_p+2t_{p1})+(2t_{p2}+4t_{p5})(1+c_z)+4t_{p3}c_z,\\
  f_1&=(\epsilon_p+4t_{p2}+4t_{p3})+(2t_{p1}+8t_{p5})c_z.
 \end{align}
\end{subequations}
On the other hand, $\hat{\mathcal{E}}'^{dd}_{\bm{k}}$ can
be explicitly written as
\begin{equation}
 \hat{\mathcal{E}}'^{dd}_{\bm{k}}=
  \begin{pmatrix}
   g_6&0&0&0&0&0\\
   0&g_7 &g_9 &0 &0 &0\\ 
   0&g_9 &g_8 &0 &0 &0\\
   0&0&0&g_6&0&0\\
   0&0 &0 &0 &g_7 & g_9\\
   0&0 &0 &0 &g_9 & g_8
  \end{pmatrix}
  \label{Edd}
\end{equation}
with
\begin{subequations} 
 \begin{align}
  g_6&=f_3-f_5,\\
  g_7&=\frac{1}{3}(f_3+2f_4+f_5-4f_6),\\
  g_8&=\frac{1}{3}(2f_3+f_4+2f_5+4f_6),\\
  g_9&=\frac{\sqrt{2}}{3}(f_3-f_4+f_5-f_6),
 \end{align}
\end{subequations}
and
\begin{subequations} 
 \begin{align}
  f_3&=(\epsilon_d+2t_{d1}+2t_{d2})+2t_{d2}c_z,\\
  f_4&=(\epsilon_d+4t_{d2})+2t_{d1}c_z,\\
  f_5&=4(t_{d0}+2t_{d3}c_z),\\
  f_6&=4(t_{d0}+2t_{d3})c_{\frac{z}{2}}.
 \end{align}
\end{subequations}
The off-diagonal matrix, $\hat{\mathcal{E}}'^{dp}_{\bm{k}}$, is
expressed as
\begin{equation}
 \hat{\mathcal{E}}'^{dp}_{\bm{k}}=
  \setlength{\arraycolsep}{0.5mm}
  \begin{pmatrix}
   -h_1k_-&h_2k_+&h_3k_+&0&0&0\\
   -h_4&0 &0 &-h'_1k_- &h'_2k_+ & h'_3k_+\\
   -h'_1k_+&-h'_2k_- &-h'_3k_- &-h_4 &0 &0\\ 
   0&0 &0 &-h_1k_+ &-h_2k_- & -h_3k_-\\
   -h'_4&0 &0 &h''_1k_- &-h''_2k_+ &-h''_3k_+\\ 
   h''_1k_+&-h''_2k_- &-h''_3k_- &h'_4 &0 &0 
  \end{pmatrix},
  \label{Edp}
\end{equation}
with 
$h_1=\mathrm{i}t_{dp}c_{\frac{z}{2}}$, 
$h_2=\mathrm{i}t_{dp}(2+c_{\frac{z}{2}})/\sqrt{3}$, 
$h_3=-2\mathrm{i}t_{dp}(1-c_{\frac{z}{2}})/\sqrt{6}$, and
$h_4=8\mathrm{i}t_{dp}s_{\frac{z}{2}}/\sqrt{3}$, 
where $h'_{1,2,3}=h_{1,2,3}/\sqrt{3}$, 
$h''_{1,2,3}=\sqrt{2}h'_{1,2,3}$, and $h'_4=h_4/\sqrt{2}$.

In the second step, we keep only the matrix elements related to
$|p_{\frac{3}{2}\frac{3}{2}}\rangle$,
$|p_{\frac{3}{2}\bar{\frac{3}{2}}}\rangle$, $|d'_2\uparrow\rangle$, and
$|d'_2\downarrow\rangle$, and drop all the other matrix elements. As a
result,
the Hamiltonian is transformed to a $4\times 4$ matrix
$\hat{\mathcal{E}}''_{\bm{k}}$, which can be written as
\begin{equation}
 \hat{\mathcal{E}}''_{\bm{k}}=
   \setlength{\arraycolsep}{1mm}
  \begin{pmatrix}
   g_1&0&h_2k_+&0\\
   0&g_1 &0 &-h_2k_-\\ 
   h_2^*k_-&0 &g_7 &0\\ 
   0&-h_2^*k_+ &0 &g_7
  \end{pmatrix}.\label{Edashdash}
\end{equation}
Finally, suppose that $g_1=g_7$ is satisfied at some $k_z=k_0$, and
expand $g_1$ and $g_7$ as
\begin{equation}
 g_1=-c_p\delta k_z+\epsilon_0,\quad
  g_7=c_d\delta k_z+\epsilon_0
  \label{diagonal_expanded}
\end{equation}
where $\delta k_z=k_z-k_{0}$ and $\epsilon_0=g_1$
($=g_7$) at $k_z=k_{0}$.
Substituting eq.~\eqref{diagonal_expanded} into eq.~\eqref{Edashdash}, the
Hamiltonian becomes,
\begin{equation}
\begin{split}
  &\hat{\mathcal{E}}'''_{\bm{k}}=\epsilon_0\hat{1}+
   \setlength{\arraycolsep}{1mm}
  \begin{pmatrix}
   -c_p\delta k_z&0&-h_2k_+&0\\
   0&-c_p\delta k_z &0 & h_2k_-\\
   -h_2^*k_-&0 &c_d\delta k_z &0\\ 
   0& h_2^*k_+&0 & c_d\delta k_z
  \end{pmatrix}\\
 &=(\epsilon_0+\delta c\delta k_z)\hat{1}+
   \setlength{\arraycolsep}{1mm}
 \begin{pmatrix}
  -c\delta k_z&0&-h_2k_+&0\\
  0&-c\delta k_z &0 & h_2k_-\\
  -h_2^*k_-&0 &c\delta k_z &0\\ 
  0&h_2^*k_+ &0 & c\delta k_z
 \end{pmatrix},
\end{split}\label{Edashdashdash}
\end{equation}
where $c$ and $\delta c$ are defined as $c=(c_d+c_p)/2$ and
$c=(c_d-c_p)/2$. At this point, it is easy to demonstrate that
eq.~\eqref{Edashdashdash} is really a (tilted) massless Dirac
Hamiltonian by changing the order of the rows and columns.
Here, we emphasize that the four basis states of this Dirac
Hamiltonian come from $|p_{\frac{3}{2}\frac{3}{2}}\rangle$,
$|p_{\frac{3}{2}\bar{\frac{3}{2}}}\rangle$, $|d_2'\uparrow\rangle$, and
$|d_2'\downarrow\rangle$. 

In deriving eq.~\eqref{Edashdash}, we have neglected the states
$|p_{\frac{3}{2}\frac{1}{2}}\rangle$,
$|p_{\frac{3}{2}\bar{\frac{1}{2}}}\rangle$, 
$|p_{\frac{1}{2}\frac{1}{2}}\rangle$, 
$|p_{\frac{1}{2}\bar{\frac{1}{2}}}\rangle$, and $|d'_1\sigma\rangle$. 
Here, let us discuss why we can neglect these states in detail.
Firstly, the energies of these states are away from the energy of the
Dirac point ($\epsilon_0$). For $|p_{\frac{1}{2}\frac{1}{2}}\rangle$ and 
$|p_{\frac{1}{2}\bar{\frac{1}{2}}}\rangle$, this is mainly because
$|g_3-\epsilon_0|$ is large at $k_z=k_{z0}$. For
$|p_{\frac{3}{2}\frac{1}{2}}\rangle$,
$|p_{\frac{3}{2}\bar{\frac{1}{2}}}\rangle$, and $|d'_1\sigma\rangle$,
this is due to the term $h_4$ in eq.~\eqref{Edp}. Namely, although
$|g_2-\epsilon_0|$ and $|g_6-\epsilon_0|$ at $k_z=k_{z0}$ are not so
large, the term $h_4$, which is finite even for $|k_\pm|=0$, makes
eigenenergies for $|p_{\frac{3}{2}\frac{1}{2}}\rangle$,
$|p_{\frac{3}{2}\bar{\frac{1}{2}}}\rangle$, and $|d'_1\sigma\rangle$
away from $\epsilon_0$. Secondly, we find that the matrix elements
between these
states and the states that we have kept in eq.~\eqref{Edp} are at most
in the first order in $k_\pm$. This means that the
contributions of these states to the low-energy Hamiltonian are at
least in the second order in $k_\pm$. 

On the other hand, we should be careful in neglecting
$|d'_3\sigma\rangle$, since the matrix elements
between $|d'_3\sigma\rangle$ and $|d'_2\sigma\rangle$ are finite even
for $|k_\pm|=0$ [see eq.~\eqref{Edp}]. In order to
eliminate these constant
off-diagonal matrix elements, we have to choose an appropriate linear
combination of
$|d'_2\sigma\rangle$ and $|d'_3\sigma\rangle$ instead of pure
$|d'_2\sigma\rangle$ as a basis state. However, using
such a linear combination only results in a renormalization of
$h_{2,3}$, $h'_{2,3}$, and $h''_{2,3}$ in eq.~\eqref{Edp} since the
transformation leading to the linear combination of $|d'_2\sigma\rangle$
and $|d'_3\sigma\rangle$ only mixes the second and the third columns (or
the fifth and the sixth columns) of the matrix in eq.~\eqref{Edp}. 
As a result, the low-energy effective Hamiltonian should have the same
form as eq.~\eqref{Edashdash}. Furthermore, the weight
of $|d'_3\sigma\rangle$ mixed in the band forming the Dirac electron is
very small as we have seen in Fig.~\ref{Fig5}(f). 

\subsection{Origin of the Dirac Electron}
As we discussed in ref.~\citen{JPSJ.80.083704}, the origin of the
Dirac electron in this model can be clarified by considering the effects
of the hybridization between the p- and d-bands ($t_{dp}$) on the band
structure. As we have pointed out, the top of p-bands locates above the
bottom of d-bands. Then, if the hybridization $t_{dp}$ is neglected, the
p- and d-bands should form usual hole- and electron-like Fermi surfaces
around the $\Gamma$-point,
and there appear no Dirac electrons. Although the p- and d-bands cross,
the crossing points are not isolated points in the Brillouin zone. Then,
we consider the effect of finite $t_{dp}$. In general, finite $t_{dp}$
causes the band anticrossing of the p- and d-bands, and eliminates the
crossing points between the p- and d-bands. However, in the present model,
the band anticrossing does not occur on the $\Gamma$--X
line because of the symmetry of wave functions
(ref.~\citen{JPSJ.80.083704} and see below). Consequently, the crossing
points are left on the special and
isolated points in the Brillouin zone that are located on the
$\Gamma$--X line. This means that there emerge Dirac points in this
model. Note that these arguments have some similarities
with those in ref.\citen{PhysRev.52.365}. 

\begin{figure}[htbp]
 \begin{center}
  \includegraphics[scale=1.0]{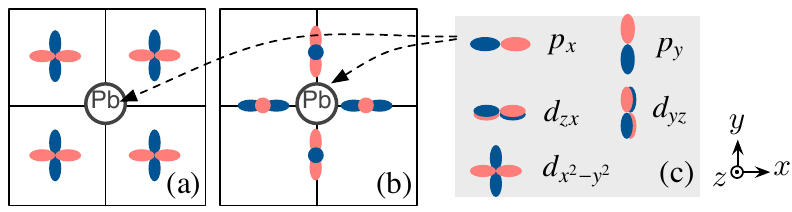}
  \caption{(Color online) (a,b) The wave functions involved in
  $|d'_2\sigma\rangle$
  states for the case that the momentum is on the $k_z$-axis. (c)
  Schematic pictures for the orbitals on the Pb site.}\label{Fig6}
 \end{center}
\end{figure}

Finally, we address why the band anticrossing does not occur on the
$\Gamma$--X line. For simplicity, we concentrate on the
Dirac point on the $k_z$-axis, which is one of the $\Gamma$--X line. In
this case, the states relevant to the Dirac electron are
$|p_{\frac{3}{2}\frac{3}{2}}\rangle$,
$|p_{\frac{3}{2}\bar{\frac{3}{2}}}\rangle$, and $|d'_2\sigma\rangle$. 
As defined in eqs.~\eqref{new_basis_p} and \eqref{new_basis_d}, 
$|p_{\frac{3}{2}\frac{3}{2}}\rangle$ and
$|p_{\frac{3}{2}\bar{\frac{3}{2}}}\rangle$ are the superpositions of
$|p_x\sigma\rangle$ and $|p_y\sigma\rangle$, while $|d'_2\sigma\rangle$
is a superposition of $|d_1\sigma\rangle$, $|d_2\sigma\rangle$, and
$|d_3\sigma\rangle$. The latter wave functions schematically depicted in 
Figs.~\ref{Fig6}(a) and \ref{Fig6}(b). From Fig.~\ref{Fig6}, it is
apparent that $|p_x\sigma\rangle$ (or $|p_y\sigma\rangle$) cannot
hybridize with the wave
functions of Figs.~\ref{Fig6}(a) and \ref{Fig6}(b)
by a symmetrical reason. This means that there are no
matrix elements between $|p_{\frac{3}{2}\frac{3}{2}}\rangle$ (or
$|p_{\frac{3}{2}\bar{\frac{3}{2}}}\rangle$) and $|d'_2\sigma\rangle$ on
the $k_z$-axis, which explains the absence of the band
anticrossing. Note that if a momentum is off from the $k_z$-axis, there
appear finite matrix elements between
$|p_{\frac{3}{2}\frac{3}{2}}\rangle$ (or
$|p_{\frac{3}{2}\bar{\frac{3}{2}}}\rangle$) and $|d'_2\sigma\rangle$.

\subsection{Origin of the Mass Term}\label{sec5}
\begin{figure}[htbp]
 \begin{center}
  \includegraphics[scale=1.0]{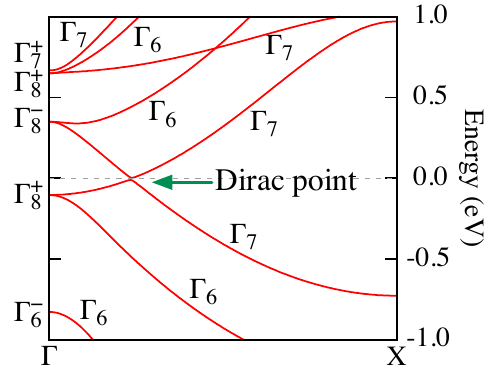}
  \caption{(Color online) Irreducible representations of the bands on
  the $\mathrm{\Gamma}$-X line. The notations used in ref.~\citen{koster}
  are used to label irreducible representations.}
  \label{Fig7}
 \end{center}
\end{figure}

The above arguments nicely explain the origin of the massless Dirac
electron in the tight-binding model. However, the first-principles
calculation shows that there exists a small but finite mass gap at the
Dirac point in Ca$_3$PbO. In this subsection, we clarify the origin of
this mass term. First, let us check the
irreducible representations to which the bands on the $\Gamma$--X line
belong. As shown in Fig.~\ref{Fig7}, it turns out that the two bands
forming the Dirac electron
belong to the same irreducible representation $\Gamma_7$.
If they belong to different representations, it is impossible to have a
mass gap. However, since they belong to the same representation, 
finite matrix elements between these two bands can exist,
which result in a finite mass gap. 

Since the mass term is very small in Ca$_3$PbO, we think that the simple
tight-binding model introduced in the previous subsection is a good
starting point discussing the mass term. In order to include the small
mass term
to the effective model, some minor effects should be included. 
In the following, we consider the effects caused by the orbitals not
included in the previous subsection. As a
representative example, we add the Pb-$5d$ orbitals
to the basis set. (The inclusion of other orbitals leads to the similar
conclusion.) Since these orbitals are located
far away from the Fermi energy (about $-15$ eV below the
Fermi energy according to the first-principles calculation), we treat
these orbitals in a perturbative way. We concentrate on the
effective model around the Dirac point on the $k_z$-axis
as before in the subsequent arguments.

The Pb-$5d$ orbitals considered here are also shown in
Fig.~\ref{Fig6}(c). From
Fig.~\ref{Fig6}(c), we can see that Pb-$d_{zx}$ (Pb-$d_{yz}$) can
hybridize with $|p_x\sigma\rangle$ ($|p_y\sigma\rangle$) in the
$z$-direction if $k_z$ is finite. As a result,
$|p_{\frac{3}{2}\frac{3}{2}}\rangle$, and 
$|p_{\frac{3}{2}\bar{\frac{3}{2}}}\rangle$, which are relevant to the
Dirac electron, can hybridize with
Pb-$d_{zx}$ and Pb-$d_{yz}$, since they are composed of
$|p_x\sigma\rangle$ and $|p_y\sigma\rangle$. 
On the other hand, we can see that Pb-$d_{x^2-y^2}$ orbital can
hybridize with the wave functions depicted in Figs.~\ref{Fig6}(a) and
\ref{Fig6}(b). As a result, $|d'_2\sigma\rangle$, which is the other
state
relevant to the Dirac electron and is composed of wave functions
in Figs.~\ref{Fig6}(a) and \ref{Fig6}(b), can hybridize with
Pb-$d_{x^2-y^2}$ orbital. To summarize, 
$|p_{\frac{3}{2}\frac{3}{2}}\rangle$,
$|p_{\frac{3}{2}\bar{\frac{3}{2}}}\rangle$, and $|d'_2\sigma\rangle$
will be modified in a perturbative way as 
\begin{align}
 |p_{\frac{3}{2}\frac{3}{2}}\rangle&\longrightarrow
 |p_{\frac{3}{2}\frac{3}{2}}\rangle
 +\frac{w_1}{\Delta_1}|\text{Pb-}d_{\frac{3}{2}}\rangle\\
 |p_{\frac{3}{2}\bar{\frac{3}{2}}}\rangle&\longrightarrow
 |p_{\frac{3}{2}\bar{\frac{3}{2}}}\rangle
 +\frac{w_1}{\Delta_1}|\text{Pb-}d_{\bar{\frac{3}{2}}}\rangle\\
 |d'_2\sigma\rangle&\longrightarrow
  |d'_2\sigma\rangle+\frac{w_2}{\Delta_2}|\text{Pb-}d_{x^2-y^2}\sigma\rangle,\label{ddash2}
\end{align}
where
\begin{align}
 |\text{Pb-}d_{\frac{3}{2}}\rangle&=-\frac{1}{\sqrt{2}}
  \Bigl(
 |\text{Pb-}d_{zx}\uparrow\rangle+\mathrm{i}|\text{Pb-}d_{yz}\uparrow\rangle
 \Bigr),\\
  |\text{Pb-}d_{\bar{\frac{3}{2}}}\rangle&=\frac{1}{\sqrt{2}}
  \Bigl(
 |\text{Pb-}d_{zx}\downarrow\rangle-\mathrm{i}|\text{Pb-}d_{yz}\downarrow\rangle
 \Bigr).
\end{align}
Here, $w_{1,2}$ and $\Delta_{1,2}$ represent the matrix elements and the
level differences between the originally included and newly added
orbitals, respectively. 
Note that $w_1$ and $w_2$ depend on $k_z$, and $w_1$ is zero at 
$k_z=0$ while $w_2$ is finite at $k_z=0$ from the symmetry of the
orbitals.

Now, assuming the standard on-site $\bm{L}\cdot\bm{S}$ coupling for Pb
$5d$ orbitals (not Pb $6p$ orbitals), there appear new matrix elements 
between $|p_{\frac{3}{2}\frac{3}{2}}\rangle$,
$|p_{\frac{3}{2}\bar{\frac{3}{2}}}\rangle$, and $|d'_2\sigma\rangle$
orbitals. Explicitly, we find that
\begin{subequations} 
 \begin{align}
 \langle
  d'_2\uparrow|\hat{H}_{\text{LS}}|p_{\frac{3}{2}\frac{3}{2}}\rangle
  &\sim
  \alpha\langle\text{Pb-}d_{x^2-y^2}\uparrow|
  \hat{H}_{\text{LS}}|\text{Pb-}d_{\frac{3}{2}}\rangle
 =0,\\
  \langle
  d'_2\uparrow|\hat{H}_{\text{LS}}|p_{\frac{3}{2}\bar{\frac{3}{2}}}\rangle
  &\sim
  \alpha\langle\text{Pb-}d_{x^2-y^2}\uparrow|
  \hat{H}_{\text{LS}}
  |\text{Pb-}d_{\bar{\frac{3}{2}}}\rangle=\sqrt{2}\alpha\lambda,
 \end{align}\label{mass:matrix_elements1}\end{subequations}
where $\alpha=\frac{w_1w_2}{\Delta_1\Delta_2}$. Similarly, we have
$\langle
d'_2\downarrow|\hat{H}_{\text{LS}}|p_{\frac{3}{2}\frac{3}{2}}\rangle
\sim \sqrt{2}\alpha\lambda$
and 
$\langle
d'_2\downarrow|\hat{H}_{\text{LS}}|p_{\frac{3}{2}\bar{\frac{3}{2}}}\rangle
=0$. 
Then, adding these contributions to eq.~\eqref{Edashdashdash}, the
effective Hamiltonian becomes
\begin{equation}
 \hat{\mathcal{E}}_{\bm{k}}'''
 =(\epsilon_0+\delta c\delta k_z)\hat{1}+
   \setlength{\arraycolsep}{1mm}
 \begin{pmatrix}
  -c\delta k_z&0&-h_2k_+&m \\
  0&-c\delta k_z &m & h_2k_-\\
  -h_2^*k_-&m &c\delta k_z &0\\ 
  m&h_2^*k_+ &0 & c\delta k_z
 \end{pmatrix},\label{mass:massive_dirac}
\end{equation}
with $m\sim\sqrt{2}\alpha\lambda$. 
This is nothing but a (tilted) massive Dirac Hamiltonian. 

We can obtain the two
important features about the mass term from
the above derivation. First, the mass term is small since it requires
the inclusion of the orbitals locating far away from the Fermi
energy. Second, the mass term scales with the coefficient for the
spin--orbit coupling, $\lambda$.

We must note that any other orbitals having the same symmetry as
Pb-$d_{zx}$, $d_{yz}$, $d_{x^2-y^2}$ orbitals can be sources of the mass
term. For example, some states originated from the Ca-$3d$ orbitals
(other than $d_{x^2-y^2}$, which is included in the original basis set)
have symmetries of Pb-$d_{zx}$, and $d_{yz}$ orbitals. Thus, inclusion
of the Ca-$3d$ orbitals other than $d_{x^2-y^2}$ will also contribute
to the mass
term. Again, these orbitals are located away from the Fermi energy, and
thus, 
the mass term induced by these orbitals are also small. 
The resultant mass term scales with $\lambda$ also in this case.

\section{Topological Property and the Surface Band
 Structure}\label{sec:surface}
\subsection{Topological Property}
In this subsection, we discuss the relation of Ca$_3$PbO to the
topological insulator. As we have stated, the top of p-bands, whose wave
functions have odd parity at the $\Gamma$-point, locates above the
bottom of d-bands, whose wave functions have even parity at the
$\Gamma$-point. This kind of structure, i.e., an overlap of the two bands
with opposite parity, is called an ``inverted'' band structure, which can
leads to a topological
insulator in some
cases\cite{PhysRevB.76.045302,JPSJ.77.031007}. Furthermore, Pb atom has
a very strong spin--orbit coupling. Then, it is
tempting to assign Ca$_3$PbO as a topological insulator. In fact,
Ca$_3$PbO and its relatives are suggested as potential topological
insulators in ref.~\citen{Klintenberg:2010uq}. However, unfortunately,
Ca$_3$PbO is not a topological insulator as explained in the following. 

The $Z_2$ topological number $\nu_0$ for Ca$_3$PbO is readily evaluated since
this material has an inversion symmetry. According to
ref.~\citen{PhysRevB.76.045302}, $\nu_0$ can be obtained from the formula 
\begin{equation}
 (-1)^{\nu_0}=\prod_{i=1}^8\delta_i
  =\prod_{i=1}^8\prod_{m=1}^N\xi_{2m}(\Gamma_i),
  \label{Z2_for_inversion_symmetric_case}
\end{equation}
for the materials having an inversion symmetry. In this formula,
$\Gamma_i$ represents the time reversal invariant momenta (TRIM) in the
Brillouin zone
and $\xi_{2m}(\Gamma_i)$ represents the parity of the wave function for
the $2m$-th state at $\Gamma_i$. $N$ is the number of filled
bands counted neglecting the Kramers degeneracy. For the simple cubic
crystal of Ca$_3$PbO, TRIM are the $\Gamma$-, X-, M-, and R-points where
there are three X- and M-points. Using WIEN2k package, we derive
the irreducible representations and the parities of the states at
TRIM. Here, we
concentrate on the states whose energies are in between $-4$ eV and $0$
eV, since the bands below $-4$ eV are associated with the completely
filled shells of the involved atoms such as O-$2p$ orbitals and give
only trivial contributions to $\nu_0$. 
There are three bands between $-4$ eV and $0$ eV, and thus, $N=3$.
The obtained results are
summarized in Table~\ref{parities}. From the table, we can see
that the right hand side of eq.~\eqref{Z2_for_inversion_symmetric_case}
is equal to $1$, which implies that this material is topologically
trivial. 
\begin{table}[htbp]
 \begin{center}
  \caption{Irreducible representations and the parities of the states
  whose energies are in between $-4$ eV and $0$ eV at TRIM.}\label{parities}
  \begin{tabular}{|c|ccc||c|c|c||c|}
   \hline
   $\Gamma_i$ &\multicolumn{3}{c||}{irreducible
   rep.}&$\xi_2$&$\xi_4$&$\xi_6$& $\delta_i$\\ \hline
   $\Gamma$ ($\times$1)&$\Gamma_6^-$ & \multicolumn{2}{c||}{$\Gamma_8^+$}&
   $-$&$+$&$+$&$-$\\
   X ($\times$3)& $\Gamma_6^-$&$\Gamma_6^-$ &$\Gamma_7^-$&$-$&$-$&$-$&$-$\\
   M ($\times$3)& $\Gamma_6^-$&$\Gamma_7^-$ &$\Gamma_6^-$&$-$&$-$&$-$&$-$\\
   R ($\times$1)&
       $\Gamma_6^-$&\multicolumn{2}{c||}{$\Gamma_8^-$}&$-$&$-$&$-$&$-$\\
   \hline
  \end{tabular}
 \end{center}
\end{table}

We can understand why Ca$_3$PbO is topologically trivial as follows.
As we have checked, there exists an
``inverted'' band structure at the $\Gamma$-point. However,
the inverted bands at the $\Gamma$-point have a double degeneracy
{\it other than} Kramers degeneracy. In other words, they have a
quadruple degeneracy if the Kramers degeneracy is included. As a
result, this ``inverted''
band structure gives $(-1)^2=1$ in the right hand side of
eq.~\eqref{Z2_for_inversion_symmetric_case}. This makes Ca$_3$PbO
topologically trivial. In order to turn this material
into a topological insulator, at least the degeneracy at the
$\Gamma$-point should be lifted by lowering the symmetry of the crystal.

\subsection{Surface Band Structure}
\begin{figure}[htbp]
 \begin{center}
  \includegraphics[scale=1.0]{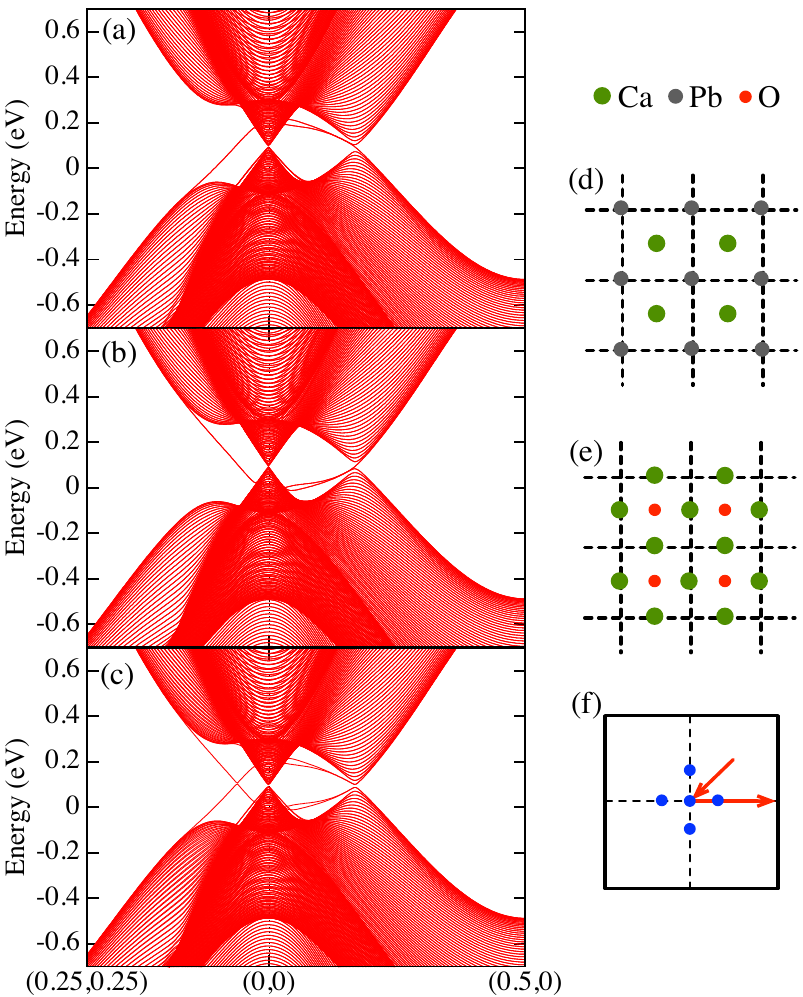}
  \caption{(Color online) Surface band structures for the 001
  surfaces. The top and bottom surfaces of the slab are (a) 001A, (b)
  001B, and (c) 001A and 001B. (d,e) Schematic pictures for (d) 001A and
  (e) 001B surfaces. (f) Surface Brillouin zone for the 001 surfaces. Dots
  represent the Dirac points projected to the surface Brillouin
  zone. Allows represent the path on which the band structure is
  plotted.}
 \label{Fig8}
 \end{center}
\end{figure}

\begin{figure}[htbp]
 \begin{center}
    \includegraphics[scale=1.0]{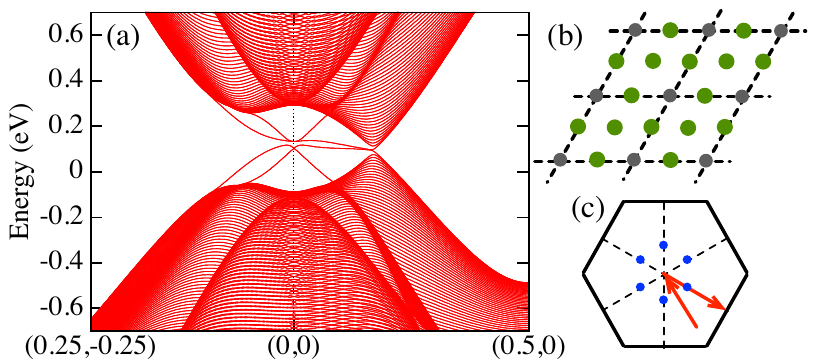}
  \caption{(Color online) (a) Surface band structure for the 111
  surface. (b) Schematic
  picture for the 111 surface. (c) Surface Brillouin zone for the 111
  surface. Dots and allows represent the same things as in
  Fig.~\ref{Fig8}.}
  \label{Fig9}
 \end{center}
\end{figure}

Although we have proved that Ca$_3$PbO is not a topological insulator,
we study the surface
band structure of Ca$_3$PbO. The surface band structures are calculated
using the
tight-binding model in a slablike lattice structure having two (top and
bottom) surfaces. Here, surfaces are
treated as simple terminations of the lattice points for simplicity
(i.e., we use open boundary conditions).
In the following, we
consider the cases in which the slab has 001 or 111
surface. In Ca$_3$PbO, there are two types of 001
surfaces. The schematic pictures for these two 001 surfaces are shown in
Fig.~\ref{Fig8}(c) and \ref{Fig8}(d). These two types of
surfaces are named as 001A [Fig.~\ref{Fig8}(c)] and
001B [Fig.~\ref{Fig8}(d)], respectively. As we can see from
Figs.~\ref{Fig8}(c) and \ref{Fig8}(d), 001A
surface has equal number of Ca and Pb sites, while 001B surface has only
Ca sites. (In Fig.~\ref{Fig8}(d), O sites are also written, but O
sites are not included in the basis set of the tight-binding model.) 
In contrast, there is only one type of 111 surface
[Fig.~\ref{Fig9}(b)]. Actually, there is another possible 111
surface, but that surface has only O sites and is equivalent to the 111
surface in Fig.~\ref{Fig9}(b).

The surface band structures for the 001 surfaces are summarized in
Figs.~\ref{Fig8}(a)--\ref{Fig8}(c). The top and bottom
surfaces of the slab used in the calculation are 001A for
Fig.~\ref{Fig8}(a), 001B for Fig.~\ref{Fig8}(b), and 001A
and 001B for Fig.~\ref{Fig8}(c). 
Figure~\ref{Fig8}(f) shows the surface Brillouin zone for the 001
surface and the arrows indicate the path on which the band
structures in Figs.~\ref{Fig8}(a)--\ref{Fig8}(c) are
shown. The positions of
the Dirac points projected on to the surface Brillouin zone are
represented by dots. Considering the relation between the
arrows and dots in Fig.~\ref{Fig8}(f), the shapes of the bulk
continuum (the region filled by a large number of bands) in
Figs.~\ref{Fig8}(a)--\ref{Fig8}(c) are easily understood:
the bulk continuum has a gap except near the projected Dirac points. 

Very interestingly, we can see that several bands appear 
near $(0,0)$ in all of Figs.~\ref{Fig8}(a)--\ref{Fig8}(c)
apart from the bulk continuum. 
We can see that these separated bands are definitely surface states
since they are affected by the choice of the
surfaces. When we closely investigate the surface states, we find that
they are doubly degenerate in Figs.~\ref{Fig8}(a) and
\ref{Fig8}(b). This degeneracy is a Kramers degeneracy since the
slab used in Figs.~\ref{Fig8}(a) and \ref{Fig8}(b) has
an inversion center because the top and bottom surfaces are identical. On
the other hand, the surface states in Fig.~\ref{Fig8}(c) are
nondegenerate. This is because the slab does not have an inversion
center
since the top and bottom surfaces are different for
Fig.~\ref{Fig8}(c). 
From these results, we can see that the nondegenerate surface states
exist when we look at one of the top or bottom surface.
In this sense, this model resembles to a
topological insulator, although we have checked that Ca$_3$PbO is not a
topological insulator. 
Note that it is natural for the surface bands to have no
Kramers degeneracy (if we look at one surface), since the spin--orbit
coupling is included in our model and the existence of a surface breaks
the inversion symmetry.

The surface band structure for 111 surface is shown in
Fig.~\ref{Fig9}(a). Again, we can see the 
surface states in Fig.~\ref{Fig9}(a). As in the case of the 001
surface, the surface bands are nondegenerate if we look at one of the
top or bottom surface. These results
suggest that the nondegenerate surface states in this model are rather
robust, although this system is not a topological insulator. 
It is a very interesting future
problem to clarify the relation between the topological insulators and
Ca$_3$PbO series.

\section{Summary and Discussion}\label{sec6}
In this paper, we first constructed a tight-binding model that describes
the low-energy band structure of Ca$_{\text{3}}$PbO. The three Pb-$p$
orbitals and three
Ca-$d_{x^2-y^2}$ orbitals are included in the basis set, reflecting the
result of the analysis of the orbital weight distributions on the
bands. The appropriate choices of the hopping integrals lead to the
satisfactory agreement between the band structure obtained in the
first-principles calculation and that of the constructed
tight-binding model. Especially, the Dirac electrons, which is the most
important feature of the band structure of Ca$_{\text{3}}$PbO, emerge in
the tight-binding model. 

It was then confirmed that the low-energy effective Hamiltonian in the
vicinity of the Fermi energy is really a Dirac Hamiltonian by applying
the basis transformation and the expansion of the matrix elements with
respect to $k_\pm$. In this procedure, four basis wave functions required
for the realization of the Dirac Hamiltonian are identified as
$|p_{\frac{3}{2}\frac{3}{2}}\rangle$,
$|p_{\frac{3}{2}\bar{\frac{3}{2}}}\rangle$, $|d'_2\uparrow\rangle$, and
$|d'_2\downarrow\rangle$. Each basis wave function corresponds
to each components of the pseudospin of the Dirac electron. Therefore the
information of the basis wave functions clarified in this paper will play
important roles in considering the possible scheme for controlling the
internal degrees of freedom of the emergent Dirac electrons.

We have also made a detailed argument on the origin of the mass term of
the Dirac electron. The conclusion is that the orbitals other than
Pb-$p$ and Ca-$d_{x^2-y^2}$ orbitals and the spin--orbit
coupling play key roles in providing the mass term. From this
analysis, we can easily understand the reason for
the smallness of the mass term. Namely, the finite mass term requires
the inclusion of the orbitals whose weights are lying far away from the
Fermi energy. Since the spin--orbit coupling on the Pb atom is
essential for the mass term, the mass term can be controlled by
replacing Pb atoms by, for example, Sn. Specifically, the alloy of
Ca$_{\text{3}}$(Pb$_{1-x}$Sn$_x$)O will give a smaller mass gap since
the spin--orbit coupling of Sn is weaker than that of Pb.
In fact, the first-principles calculation for Ca$_{\text{3}}$SnO
results in the smaller mass gap compared with that obtained in
Ca$_{\text{3}}$PbO\cite{JPSJ.80.083704}. 

Finally, we have discussed the surface band
structures of Ca$_3$PbO using the constructed tight-binding model. It is
found that there exist nontrivial surface
bands that cannot be explained as the bulk states projected on the
surface Brillouin zone. Note that
ref.~\citen{PhysRevB.83.205101} shows
that there should be nontrivial surface states for materials having Weyl
fermions. Interestingly, the observed surface states are nondegenerate
if we look at one surface. Further studies on the surface band
structure in the present model, such as to investigate the relation
between Ca$_3$PbO and topological insulators, are highly desired. 
In this paper surfaces were treated as simple open
boundaries. When we compare the present results and some future
experimental results, it
will be required to consider the effect of surface potential since the
surfaces in Figs.~\ref{Fig8}(c), \ref{Fig8}(d), and
\ref{Fig9}(b) are not charge neutral.

The successful construction of the simple tight-binding model for
Ca$_{\text{3}}$PbO encourages further search for new materials having
Dirac electrons.
The detailed analysis provided in
this paper are useful in the future experimental and theoretical
studies of the Ca$_{\text{3}}$PbO and related materials. 

\begin{acknowledgments}
T.K. was supported by JSPS Research Fellowship. 
\end{acknowledgments}

\end{document}